# Atom-by-atom extraction using scanning tunneling microscope tip-cluster interaction

***A. Deshpande[1], H. Yildirim[2,3], A. Kara[2,3], D. P. Acharya[1], J. Vaughn[1], T. S. Rahman[2,3], S.-W. Hla[1,*]***

[1]*Physics & Astronomy Department, Ohio University, Athens, OH 45701.*
[2]*Department of Physics, Cardwell Hall 116, Kansas State University, Manhattan, Kansas 66506*
[3]*Department of Physics, University of Central Florida, Orlando, Florida 32816*

**We investigate atomistic details of a single atom extraction process realized by using scanning tunneling microscope (STM) tip-cluster interaction on a Ag(111) surface at 6 K. Single atoms are extracted from a silver cluster one atom at a time using small tunneling biases less than 35 mV and a threshold tunneling resistance of 47 ± 2 kΩ. A combination of total energy calculations and molecular dynamics simulations shows a lowering of the atom extraction barrier upon approaching the tip to the cluster. Thus, a mere tuning of the proximity between the tip and the cluster governs the extraction process and is sufficient to extract an atom. The atomically precise control and reproducibility of the process are demonstrated by repeatedly extracting single atoms from a silver cluster on an atom-by-atom basis.**

[*]Corresponding author, Email: hla@ohio.edu, web: www.phy.ohiou.edu/~hla

PACS: 82.37.Gk, 68.43.-h, 81.07.-b, 68.37.Ef, 33.15.Bh

The interaction between a scanning probe tip and an adsorbate continues to unveil diverse phenomena in surface physics. In recent years, the manipulation of single atoms and molecules has been a major advance in the application of the scanning tunneling microscope (STM) [1-18]. The main appeal of STM manipulation is the ability to access, control and modify the interactions between the tip and the adsorbate, a few angstroms apart [2-6, 9]. A STM manipulation procedure known as lateral manipulation (LM) [2,9] utilizes the tip-atom interactions to precisely position individual atoms on a surface thereby allowing the construction of artificial nanostructures on an atom-by-atom basis [1,3,4,6,7]. The STM tip-height or current signals during LM provide information about the dynamical events responsible for the phenomenon, revealing the intrinsic details of the tip-atom interactions [2,6,17,18]. To date atom manipulation using a STM or an AFM -tip [19] has been restricted to flat surfaces. In this letter, we report extraction and manipulation of individual atoms on three dimensional nanoclusters on a Ag(111) surface. The atomistic details of the atom extraction mechanism are explained by means of statistical analyses and theoretical modeling, which reveals that just by locating the STM-tip at required proximity of the nanocluster greatly reduces the extraction barrier facilitating repeated removal of the top atoms from the cluster.

The experiment was performed using a homebuilt ultra-high-vacuum, low-temperature STM (UHV-LT-STM) operated at 6K [20]. The Ag(111) sample was cleaned by repeated sputter-anneal cycles. An electrochemically etched polycrystalline tungsten wire was used as the STM tip. The tip was prepared insitu by dipping into the silver substrate prior to the experiment [3]. This procedure reshapes the tip-apex and makes it atomically sharp. Moreover, the tip gets coated with the substrate material, i.e. silver atoms, and thus, the chemical identity of the tip-apex is known [3]. Such an atomically sharp and chemically identified tip-apex is crucial for a reproducible atom extraction process.

Once the desired quality and shape of the tip is obtained, a large terrace on the Ag(111) surface is selected for the manipulation experiment. A nanometer size silver cluster is then deposited on the surface by making a controlled tip-sample contact (Fig. 1). The landscape of the cluster is carefully analyzed from collected STM images. The irregularities in the shape of the cluster, seen as protrusions in three dimensional image processing, are chosen to be the ideal target zones for the tip to extract the atoms with ease (Fig. 1a).

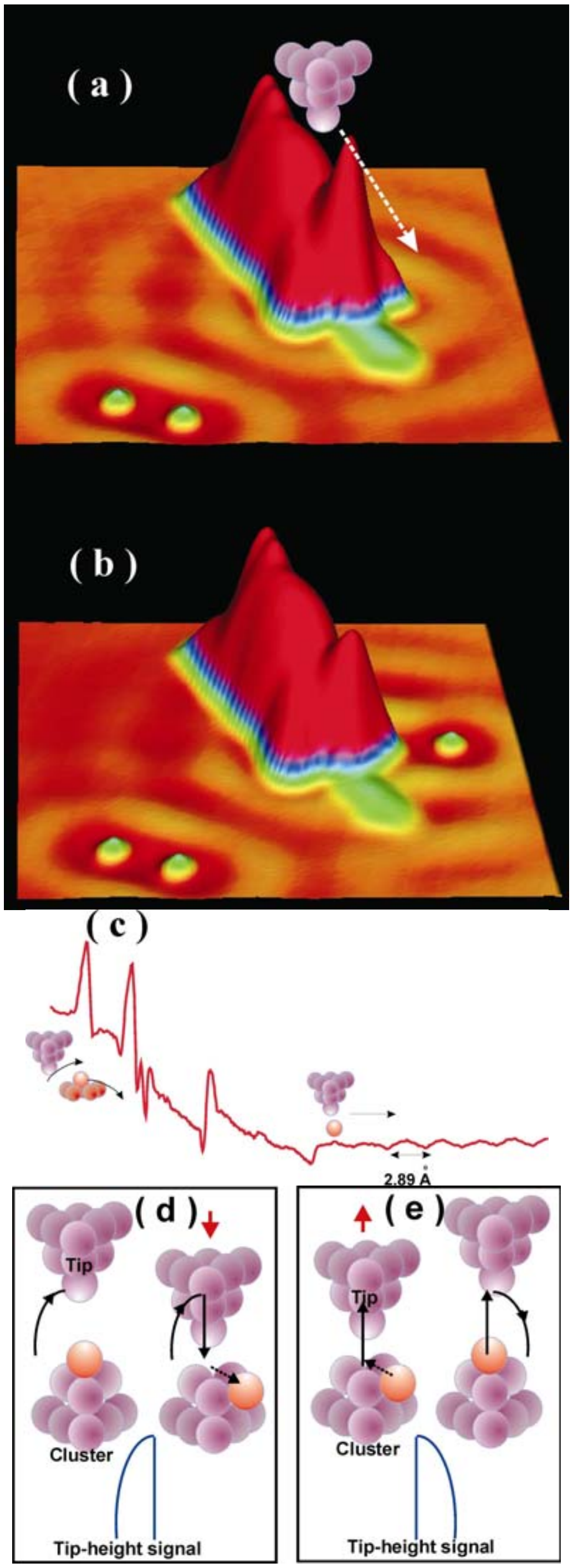

***Figure 1.*** *Atom extraction. (**a**) A three-dimensional STM image of a silver nanocluster deposited by tip-surface contact. The tip is brought close to the protruded part of the cluster and then moved laterally towards a destination on the Ag(111) surface. (**b**) An STM image acquired after this procedure shows a height reduction of the cluster protrusion and the extracted atom at the final surface destination. (**c**) The manipulation signal of this event reveals the atomistic*

*details of the atom extraction. High peaks at the left side are caused by removal of the atom from the favorable adsorption sites on the cluster. The smooth single atom periodicity at the right side of the curve is due to the sliding mode manipulation of extracted atom on the flat terrace. (**d**) The drawings demonstrate the tip climbing up along the contour of top atom inside the cluster (left) and an abrupt decrease in tip height due to the removal of atom from the site (right). (**e**) A reverse process of (d) where the atom moves under the tip causes an abrupt increase in tip height followed by the tip moving a part of down slope of the atom.*

Figure 1a and 1b show STM images before and after an atom extraction process. Figure 1c shows the manipulation signal recorded during an atom extraction along with an illustration as a guide to the atomistic process. For the extraction, the tip is initially positioned near a protrusion of the cluster and the tip-height is then reduced. An increased tip-cluster interaction is thus achieved. The tip is laterally moved from one side of the cluster towards the terrace for a short distance using the constant current mode of manipulation. The subsequent STM image taken after the manipulation confirms the successful atom-extraction (Fig. 1b). A range of bias voltages from 10mV to 35 mV was used in this scheme.

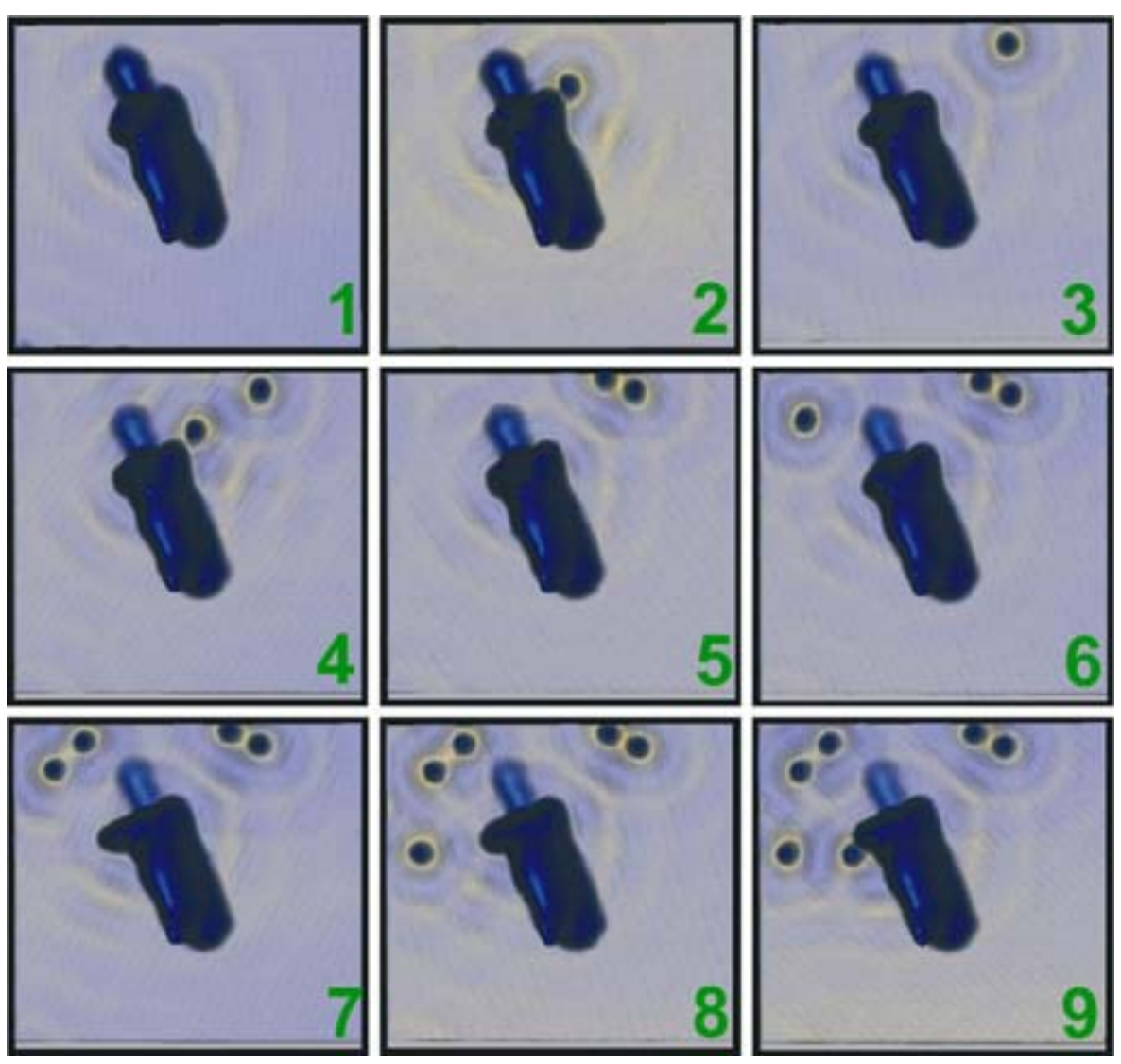

***Figure 2.*** *Atom-by-atom extraction sequence. A sequence of STM images from an STM movie [23] shows atom-by-atom extraction from a silver cluster. This series proves the reliability of our atom extraction process.*

The dynamics of this atom extraction process can be inferred from the STM manipulation signals (tip height curves). An example is illustrated in Fig. 1c the atomistic behavior during this process can be explained as follows: The manipulation signal shows a peak at the topmost location of the cluster caused by sudden changes of vertical tip position. Over hundreds of repeated atom extraction trials confirm that such peaks can be associated with a successful atom extraction. There are two types of tip-height peaks: the first type, which is commonly observed at the initial atom extraction, shows an upward curve followed by a sharp drop while the second type includes a sharp increase followed by a downward curve. Fig. 1d explains the mechanism for the first peak. Initially, the tip climes up the contour of protruding part producing a curve-like increase in tip-height. When the topmost atom of the cluster is moved away from its original position, the tip-height drops suddenly because of the constant current scanning mode. Since the extracted atom is a constituent of the cluster, removing it from the original position involves severing the bonds with neighboring atoms within the cluster. This event leads to re-adsorption of the atom at the next favorable site at the downward slope of the cluster. Several subsequent atom removals occur along the lateral tip movement providing more peaks in the manipulation signal. If the atom is loosely bound at the next site, it can easily move back under the tip. This would cause an abrupt increase in tip-height followed by downward slope of the tip along the atomic contour (second type peak, Fig. 1e). At the flat terrace, the tip height suddenly increases followed by a smooth manipulation curve. This event is associated with the positioning of the extracted atom under the tip apex. Following this, the manipulation mode is changed to that of “sliding” [2,17,18,20]. In a sliding mode, the manipulated atom is temporarily bound to or trapped under the tip while still remaining on the surface, and both the tip and the atom move together smoothly across the surface [2,17,18,20]. This behavior is consistently observed throughout the atom extraction process [23]. In order to demonstrate the reliability of our atom extraction process, we present a series of snapshot images from an STM movie (Fig. 2) (watch the movie in EPAS document [23]) that show repeated atom-by-atom extraction from a silver cluster using the procedure described above.

Even though the atomistic dynamic of atom extraction can be understood from the already established knowledge of manipulation signals [17,18,20], the environment that the extracted atom faces during the process is clearly different from the atom manipulation on a flat surface. In particular, this atom extraction involves pulling out the top atom from a protruding part of a cluster, and then moving it along a rough terrain on a three dimensional cluster surface. This motivates us to investigate further the underlying process in details. In order to gain a deeper insight of the mechanism, we first determine the threshold tunneling resistance ($R_{th}$) necessary to extract an atom from a cluster. The probability of atom extraction is defined as the ratio between the distance traveled by the atom ($L_{atom}$) and the lateral tip movement distance ($L_{tip}$). The $L_{atom}/L_{tip}$ values for atom extraction events are determined by varying the tunneling current at different fixed tunneling biases. A plot of $\boldsymbol{L_{atom}/L_{tip}}$ as a function of tunneling current measured at a fixed 18 mV bias is presented in Fig. 3a. Tunneling resistance is a qualitative measure of both, the tip-cluster distance and the tip-cluster interaction strength. For example, lowering the tunneling resistance causes a reduced tip-cluster distance and consequently an increased tip-cluster interaction. The $\boldsymbol{L_{atom}/L_{tip}}$ is zero below 400 nA tunneling current, and thus no atom has been extracted below that value (Fig. 3a). The $\boldsymbol{L_{atom}/L_{tip}}$ is ~1 above 400 nA implying that the atom can be successfully extracted. In a few cases, $\boldsymbol{L_{atom}/L_{tip}}$ remains zero even at higher current (>400nA) indicating that not all extraction events are successful. We find a success rate of ~85% from all the trials in the experiment. The plot in Fig. 3a clearly has a characteristic of a step-function with a current threshold at 400nA. Next, we repeat the same procedure by changing different biases and the results are displayed in Fig. 3b as a plot of threshold current versus bias. Here, each data point is determined by plotting the curves as shown in Fig. 3a. From Fig. 3b, the $\boldsymbol{R_{th}}$ value of 47 ± 2 kΩ is determined from the slope of the curve, which can be related to the threshold tip-cluster distance for atom extraction via a tip-height versus tunneling resistance curve (Fig. 3c) [21]. Here, the threshold tip-cluster distance corresponding to $\boldsymbol{R_{th}}$ = 47 ± 2 kΩ is determined to be 0.6 Å (Fig. 3c). This means that ≤ 0.6 Å distance between the edges of van der Waals radii of the tip-apex atom, and the cluster atom is necessary for a successful atom extraction (Fig. 3d). For the manipulation of individual silver atoms on a flat Ag(111) terrace, $\boldsymbol{R_{th}}$ = 210 ± 19 kΩ corresponding to a tip-atom distance of 1.3 ± 0.2 Å is required [6].

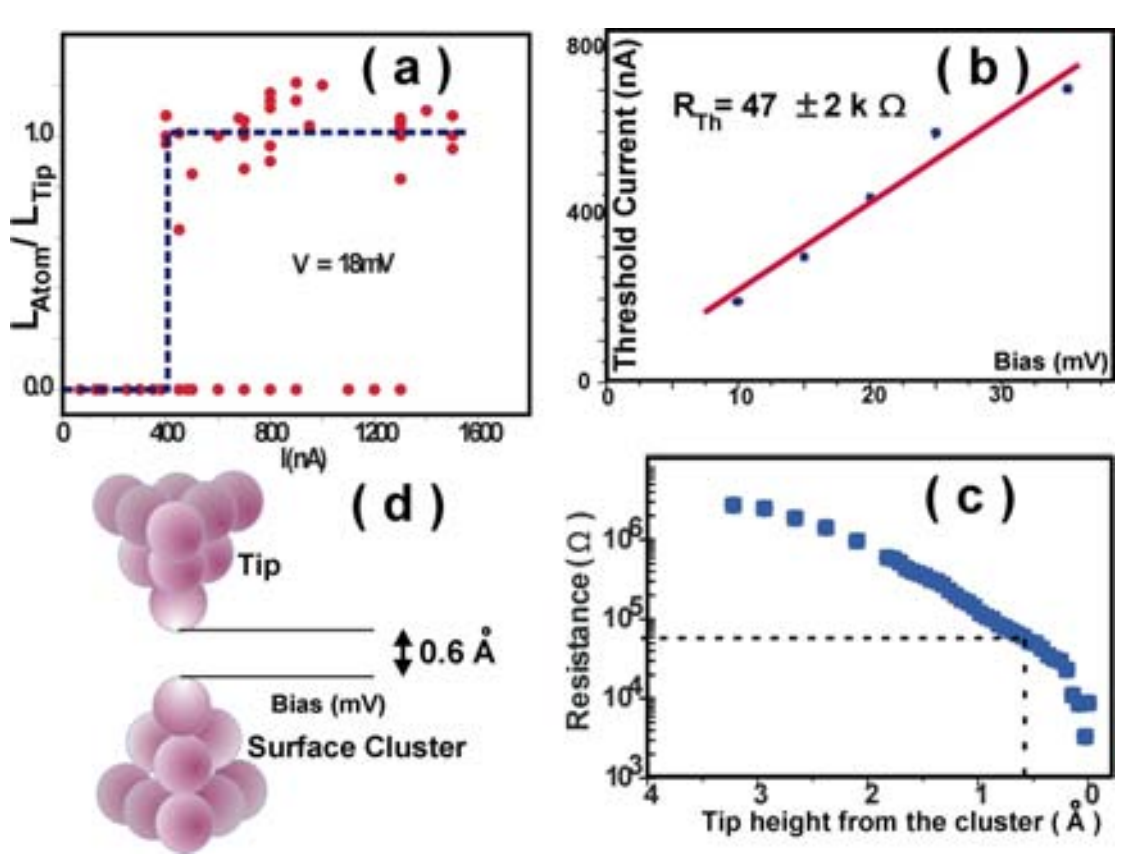

***Figure 3.*** *Threshold resistance and tip-height measurements. (**a**) A plot of* $\boldsymbol{L_{atom}/L_{tip}}$ *versus tunneling current measured at a fixed tunneling bias of 18 mV shows a characteristic step function at 400nA. (**b**) The threshold tunneling current versus bias plot shows a linear relationship. The threshold tunneling resistance* $\boldsymbol{R_{th}}$ *to extract an atom is determined from the slope of this curve.* ***(c)*** *Tunneling resistance versus tip-height plot. Here the tip-cluster distance of 0.6 Å corresponding to* $\boldsymbol{R_{th}}$ *= 47 ± 2 kΩ is obtained (shown with a dashed line). (**d**) This cartoon model demonstrates the required tip-cluster distance of 0.6 Å to extract the top-most atom of the cluster.*

Within a small bias range investigated here, the threshold tunneling current varies linearly with the bias (Fig. 3b). If local heating or tunneling electron induced excitations are involved in the process, the curve in Fig. 3b would deviate from linearity [14]. The linear relationship indicates that influence of such effects is not significant even though it can not be ruled out completely. Furthermore, since very low voltages (< 35 mV) are used, the electric field contribution in the process is negligible [3, 14]. Therefore, the tip-cluster interaction should be the central element in this case. We find that the threshold resistance (or current as shown in Fig.3a, and 3b) is a well-defined and reproducible quantity for the atom extraction as in the case of atom manipulation on a flat terrace [6] even though a huge variability in possible cluster shapes.

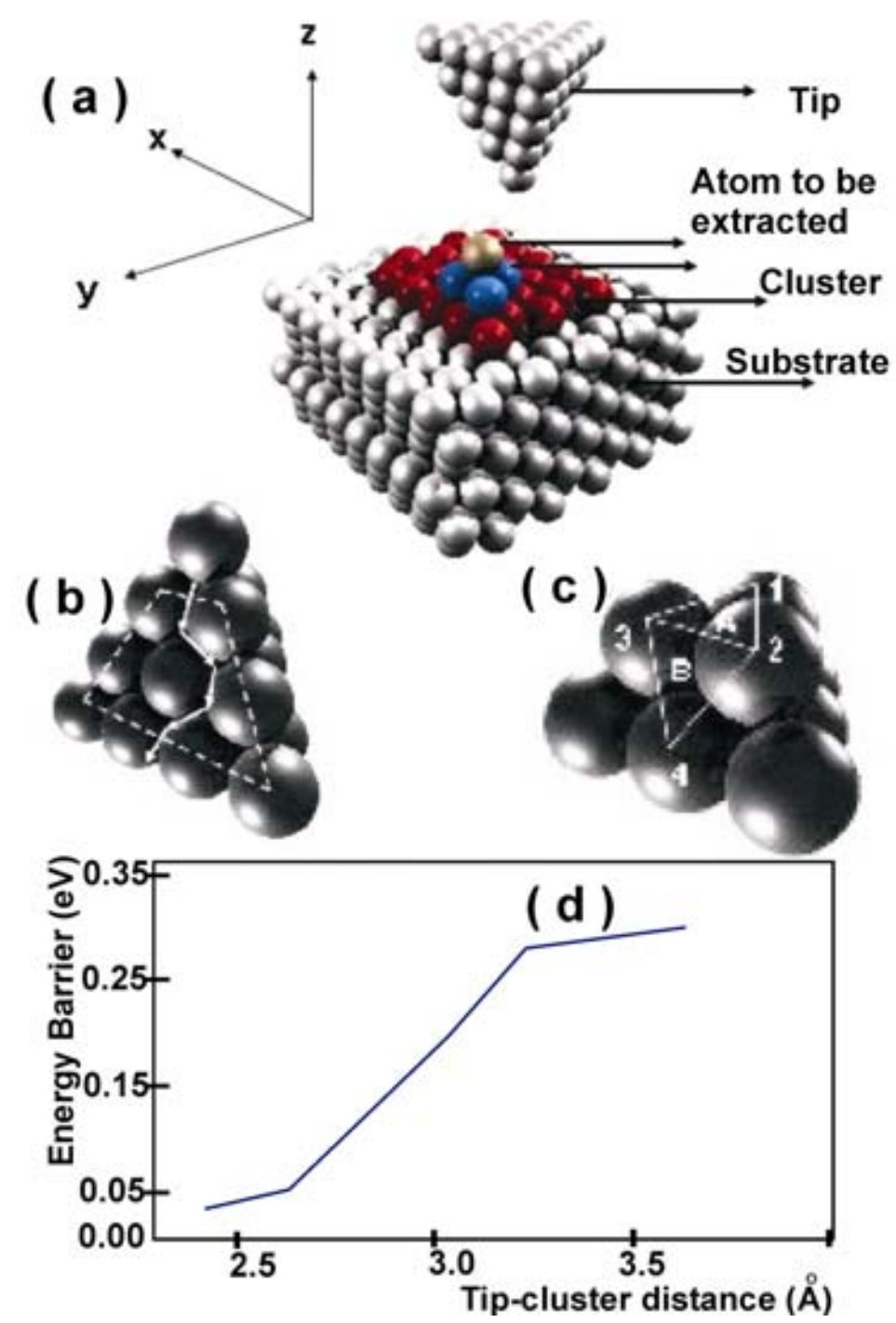

***Figure 4.*** *Theoretical modeling of the tip-cluster junction. (**a**) The geometry of the tip, adatom, cluster and silver substrate used for the calculation. (**b**) The minimum energy diffusion path of the atom after extraction with the STM-tip. (**c**) Transferring of the atom from A to B, and from B to A will cost the same energy. (**d**) The energy barrier to extract the top-most atom in the cluster as a function of the tip-cluster distance.*

The presence of the tip is expected to change the potential energy landscape surrounding the atom to be extracted. To obtain a qualitative picture of the process, a combined theoretical investigation using molecular dynamics and molecular statics simulations with interaction potentials as obtained from the Embedded Atom Method (EAM) [15,16] has been performed. The prototype system used for the calculations consists of a silver tip composed of 35 atoms, and a 28-atom silver cluster (Fig. 4a) on Ag(111). The substrate is formed by 6 atomic layers with 80 atoms in each layer. Initially, the entire system (Fig. 4a) is allowed to relax to its minimum energy configuration using the conjugate gradient method. Since only small biases are used in the experiment, the effect of the electric field is not considered. The molecular statics calculations provide the potential energy landscape of the system while the molecular dynamics simulations reveal the path of the atom during the extraction process.

The energy required to extract the atom from the cluster is the energy barrier to move the atom from its original location, A, to the next site, B, within the cluster (Fig. 4b and 4c). In the absence of the tip, the energy barrier for the atom to diffuse over the step edge is 300 meV, which is much higher than the 35 meV barrier for a silver atom diffusion on a flat Ag(111) terrace. These barriers are altered by the presence of the tip as the latter drastically modifies the energy landscape of the system [16]. The calculation reveals that the variation of the tip height has a dramatic effect on the potential energy of the cluster and of the extracted atom. It can be seen in Fig. 4d that by lowering the tip-height, the energy barrier for the atom to move from site A to B is greatly reduced. For example, at 2.43 Å tip-cluster distance, the barrier is merely 0.032 eV. On the other hand, the barrier for the atom to switch back from B to A, for the same tip height, is 0.180 eV [22]. Thus, we conclude that the location of the tip in close proximity of the cluster is sufficient to extract the top-atom by overcoming the binding of the atoms within the cluster [22].

In summary we have demonstrated a reproducible atom-by-atom extraction process using tip-cluster interaction. The required tip-cluster distance, 0.6 Å, for atom extraction is also experimentally determined. Theoretical calculations reveal that the location of the tip near a cluster can effectively reduce the atom extraction barrier and that the tip-cluster interaction alone is sufficient to knock out an atom from the cluster. This work not only provides a fundamental understanding of the influence of distance dependent tip-cluster interaction but also opens a novel route to produce single atoms for future nanoscale experiments or for atomistic constructions. We acknowledge thoughtful discussions with Kai-Felix Braun. This work is supported by the US Department of Energy, BES DE-FG02-02ER46012 (the OU team) and BES DE-FG03-97ER45650 (the KSU/UCF team).

**References and notes.**

(Watch the STM movie at: www.phy.ohiou.edu/~hla/atom-extract.htm)